\documentclass[%
 reprint,
 amsmath,amssymb,
 aps,%prl
pra,
]{revtex4-1}

\usepackage{graphicx}% Include figure files
\usepackage{dcolumn}% Align table columns on decimal point
\usepackage{bm}% bold math
\begin{document}

\preprint{APS/123-QED}

\title{Measurement-Device-Independent Quantum Coin Tossing}% Force line breaks with \\

\author{Liangyuan Zhao}
\author{Zhenqiang Yin}%
 \email{yinzheqi@mail.ustc.edu.cn}
\author{Shuang Wang}
\author{Wei Chen}
\email{kooky@mail.ustc.edu.cn}
\author{Hua Chen}
\author{Guangcan Guo}
\author{Zhengfu Han}
\affiliation{%
Key Laboratory of Quantum Information, University of Science and Technology of China, CAS, Hefei, Anhui 230026, China\\
Synergetic Innovation Center of Quantum Information $\&$ Quantum Physics, University of Science and Technology of China, Hefei, Anhui 230026, China
}%


%\affiliation{
%\affiliation{
 %Third institution, the second for Charlie Author
%\affiliation{%


\date{\today}% It is always \today, today,

\begin{abstract}
%An article usually includes an abstract, a concise summary of the work
%covered at length in the main body of the article.
Quantum coin tossing (QCT) is an important primitive of quantum cryptography and has received continuous interest. However, in practical QCT, Bob's detectors can be subjected to detector-side channel attacks launched by dishonest Alice, which will possibly make the protocol completely insecure. Here, we report a simple strategy of a detector-blinding attack based on a recent experiment. To remove all the detector side-channels, we present a solution of measurement-device-independent QCT (MDI-QCT). This method is similar to the idea of MDI quantum key distribution (QKD).  MDI-QCT is loss-tolerant with single-photon sources and has the same bias as the original loss-tolerant QCT under a coherent attack. Moreover, it provides the potential advantage of doubling the secure distance for some special case. Finally, MDI-QCT can also be modified to fit the weak coherent state sources. Thus, based on the rapid development of practical MDI-QKD, our proposal can be implemented easily.

\begin{description}
\item[PACS numbers]

\verb+03.67.Dd,03.67.Hk+
\end{description}
\end{abstract}

\pacs{Valid PACS appear here}% PACS, the Physics and Astronomy
\maketitle

%\tableofcontents

\section{Introduction}

Quantum coin tossing (QCT) allows two spatially separated mistrustful parties, e.g., Alice and Bob, to generate a common random bit via a quantum channel. It is the quantum counterpart of classical coin tossing first introduced by Blum \cite{blum1981coin} in 1981. QCT can be used as a primitive in many applications, such as secure multi-party computation.

According to whether the parties have preferences on the coin's output, coin tossing (CT) can be divided into two classes: weak CT (WCT) and strong CT (SCT). In WCT, the parties have opposite preferences for the coin's outcome, resulting in a winner and a loser. However, in SCT, each party's desired outcome is random. In this paper, we only refer to SCT unless otherwise noted.

The security of a QCT protocol involves three requirements: (1) If both Alice and Bob are honest, the probabilities of the coin's outcome $x\in\{0,1\}$ are $Pr[x=0] = Pr[x=1]= \frac{1}{2}$. Thus, the protocol is called \emph{correct}. (2) If Alice is dishonest and Bob is honest, the maximal probability that Alice can bias the outcome of $x$ is $Pr_{A}[x] = \frac{1}{2} + \varepsilon_{A}$. (3) If Bob is dishonest and Alice is honest, the maximal probability that Bob can bias the outcome of $x$ is $Pr_{B}[x] = \frac{1}{2} + \varepsilon_{B}$. The \emph{bias} of the protocol is defined as $\varepsilon = max \{\varepsilon_{A}, \varepsilon_{B}\}$, and the protocol is called \emph{fair} if $\varepsilon_{A} = \varepsilon_{B}$. The \emph{bias} of an \emph{ideal} QCT is 0.

The seminal QCT protocol was investigated by Bennett and Brassard in their 1984 paper \cite{bennett1984quantum}, henceforth referred to as BB84 QCT. Unfortunately, as argued in \cite{bennett1984quantum} and the following Mayers-Lo-Chau (MLC) \cite{lo1998quantum,mayers1997unconditionally,lo1997quantum} no-go theorem, an unconditionally secure \emph{ideal} QCT is impossible. However, this is not the end of the story. Based on the laws of quantum mechanics, QCT can achieve a maximal cheating probability lower than 1, which is impossible for any non-relativistic classical coin tossing unless unproven computational assumptions are made. Aharonov \emph{et al.} \cite{aharonov2000quantum} and their followers \cite{ambainis2001new,spekkens2002optimization,spekkens2001degrees,nayak2003bit,colbeck2007entanglement} indeed showed bit-commitment-based QCT protocols with $\varepsilon < \frac{1}{2}$ and achieved a temporary lower bound $\varepsilon = \frac{1}{4}$. Meanwhile, Kitaev \cite{kitaev2003quantum} proved that any \emph{fair} QCT protocol cannot achieve a \emph{bias} less than $\frac{1}{\sqrt{2}}-\frac{1}{2}\approx 0.207$. Later, the gap was closed by Chailloux and Kerenidis \cite{chailloux2009optimal}, who proposed a protocol with a \emph{bias} arbitrarily close to $\frac{1}{\sqrt{2}}-\frac{1}{2}$ based on the possibility of quantum WCT \cite{mochon2007quantum,aharonov2014simpler}.

A major flaw of the above protocols is that they only consider the perfect case. The security may be completely broken if the imperfections of practical systems are taken into account. To bridge the gap between theory and practice, several countermeasures against these imperfections have been developed. To address the problem of noise, the primitive of string coin tossing \cite{barrett2004quantum,lamoureux2005provably} was considered. H{\"a}nggi and Wullschleger \cite{hanggi2011tight} proposed that the honest players would abort with a certain probability depending on the noise level. Recently, a nested-structure framework is presented for single-short QCT and partial noise-tolerant property is achieved \cite{PhysRevA.92.022313}. Another notorious problem in long distance quantum cryptography is losses, which make the early two QCT experiments \cite{molina2005experimental,nguyen2008experimental} completely insecure. The earlier protocols in, for instance, Refs.~\cite{aharonov2000quantum,ambainis2001new}, also could not tolerate losses because of the sending nothing attack \cite{berlin2009fair} or the unambiguous discrimination (UD) and maximal-confidence discrimination (MCD) of non-orthogonal states attack\cite{bae2015quantum}.  A breakthrough was made by Berlin \emph{et al.} \cite{berlin2009fair}, who proposed a loss-tolerant QCT protocol that is completely impervious to loss of quantum states with single-photon source. We will refer to it as the BBBG09 protocol. Subsequently, BBBG09 was implemented with an entangled source because the protocol is insecure in the presence of multi-photon pulses \cite{berlin2011experimental}. Recently, by fixing the number of emitted pulses, hence limiting the probabilities of the multi-photon pulses, Pappa \emph{et al.} \cite{pappa2011practical} made the weak coherent state source applicable to the loss-tolerant protocol. However, the protocol of Pappa \emph{et al.} (referred to in the following as PCDK11 protocol) is not completely loss-tolerant. Thus, there is an upper bound of the amount of tolerated losses if one wants to achieve a security better than that of classical protocols. The PCDK11 protocol was implemented based on a commercial plug-and-play scheme designed for quantum key distribution (QKD) with several modifications \cite{pappa2014experimental}, considering all the standard realistic imperfections. The \emph{biases} of the experiments in \cite{berlin2011experimental} and \cite{pappa2014experimental} are both approximately 0.4, and they achieve distances of $10m$ and $15km$, respectively, wherein the maximal cheating probabilities are strictly lower than those classically possible.

Apart from the standard realistic imperfections, Pappa \emph{et al.} also examined how a few prevalent side channel loopholes in practical QKD affected the security of practical QCT. Although not all the attacks are effective, there remain some powerful detector-side channel attacks, e.g., detector-blinding attack, being defended. The principle of the detector-blinding attack on the plug-and-play QCT is as follows. First, a malicious Alice blinds Bob's detector, as what the eavesdropper does in the detector-blinding attack on QKD \cite{lydersen2010hacking}, and sends a honest state to Bob. Consequently, Alice can correlate her basis choice with Bob's successful detections. Then, Alice can declare any state she has sent to Bob according to Bob's random bit value and her desired coin's outcome. In this way, Alice has complete control over the coin's outcome without being detected as cheating, i.e., $\varepsilon_{A}=\frac{1}{2}$.

To remove all the known and unknown detector-side channel attacks launched by Alice in practical QCT systems, we propose a measurement-device-independent QCT (MDI-QCT) protocol based on BBBG09, which benefits from the idea of MDI-QKD \cite{lo2012measurement,xu2015measurement}. In MDI-QKD, each of the two legitimate parties sends one of the four BB84 photon states randomly to a third party, which may be the eavesdropper. The third party projects the incoming photons into a Bell state and announces the measurement result. After repeating this many times, these Bell state results can be used by the two legitimate parties to estimate the degree of eavesdropping in the post-processing. The advantage of MDI-QKD is that the legitimate parties only need to characterize their state preparations and they do not need to hold a measurement device anymore. Thus, the measurement device can be viewed as a black box and it naturally removes all the detector side-channels. The concept of MDI-QCT is similar to MDI-QKD in that Alice and Bob only need to know their state preparation processes and Bob does not have to trust his measurement device. That is to say, Bob can treat his measurement device as a black box and just obtain a Bell state outcome from it. Combining this with his and Alice's announced states, Bob can estimate whether Alice is cheating with a non-vanishing (but non-unit) probability. Thus, what would protect against the eavesdropper's attacks in MDI-QKD, is now argued to protect against dishonest Alice's detector-side channel attacks. Adapting MDI paradigm in QCT is not a quite easy issue, since Alice and Bob in QCT are mistrustful, while they are trustworthy in QKD. Note that the untrusted measurement device now is in Bob's laboratory, which is also different from MDI-QKD. We emphasize that Bob should shield his source for preventing both Alice and the untrusted measurement device from knowing the classical information about his state preparation.

In the follow of this paper, we first describe the protocol when each party uses a single-photon source in Section \ref{mdi}, and give its security analysis in section \ref{security}. Specially, in the security against a dishonest Alice, we analyze an individual attack and a coherent attack. The coherent attack comes from the conclusion of the seminal work of Ref. \cite{spekkens2002optimization}. At last, we will show the protocol is also suitable for weak coherent state source with a subtle modification.

\section{Protocol for MDI-QCT} \label{mdi}

For a QCT protocol to be partial security and practicality, there are two aspects that should be considered seriously. In theory, the protocol must circumvent the MLC no go theorem. In practice, the protocol ought to resist against the attacks due to losses. To solve these two critical problems, BBBG09 takes the BB84 template and the states in the protocol of Aharonov et al. \cite{aharonov2000quantum}, known as ATVY states $|\phi_{\alpha, a}\rangle$:
\begin{eqnarray*}
|\phi_{\alpha,0}\rangle = \sqrt{y}|0\rangle + (-1)^{\alpha}\sqrt{1-y}|1\rangle,\\
|\phi_{\alpha,1}\rangle = \sqrt{1-y}|0\rangle - (-1)^{\alpha}\sqrt{y}|1\rangle,
\end{eqnarray*}
where $y\in (\frac{1}{2},1)$, which will be adjusted to make the protocol \emph{fair}. Traditionally, $\alpha$ is the basis and $a$ is the bit.

In BBBG09, Alice partially commits either $a=0$ or $a=1$ to Bob. Thus, the density matrices $\rho_{0}=\frac{1}{2}|\phi_{0,0}\rangle\langle \phi_{0,0}|+\frac{1}{2}|\phi_{1,0}\rangle\langle \phi_{1,0}|$ and $\rho_{1}=\frac{1}{2}|\phi_{0,1}\rangle\langle \phi_{0,1}|+\frac{1}{2}|\phi_{1,1}\rangle\langle \phi_{1,1}|$ that correspond to Alice's partial commitments are distinct, which implies that the MLC no go theorem does not apply here. Moreover, Bob cannot through the UD or MCD to obtain a better discrimination of $\rho_{0}$ and $\rho_{1}$ than the minimum-error discrimination (MED). This is the key that, in BBBG09, Alice can resend the states at once if Bob's measurement fails and the protocol is loss-tolerant with single-photon source.

Note that BBBG09 protocol can work with an entangled source \cite{berlin2011experimental}. Therefore, to remove all the detector-side channel attacks launched by Alice, like MDI-QKD, we can implement the entanglement-based BBBG09 protocol in a time-reversal way and obtain the MDI-QCT. The key ingredients in MDI-QCT is the same with BBBG09, where BB84 template and ATVY states are employed. The mainly difference is that Bob will not have to trust his measurement devices via a Bell state measurement (BSM). Bob neither needs to fully characterize the BSM, nor protect it from unwanted leakage of classical information to the outside. The only classical information that Bob should protect from leakage to the untrusted measurement device and Alice is of his state preparation. First, let us show a simple example of our MDI-QCT protocol, where each party uses a single-photon source for the aim of completely tolerating losses.

\emph{Protocol: MDI-QCT}---

1. Alice picks, uniformly at random, a basis $\alpha \in\{0, 1\}$ and a bit $a \in\{0, 1\}$. She then prepares the polarization ATVY state $|\phi_{\alpha, a}\rangle$, i.e.,
    \begin{equation} \label{eq:states}
    \begin{aligned}
    |\phi_{\alpha,0}\rangle = \sqrt{y}|H\rangle + (-1)^{\alpha}\sqrt{1-y}|V\rangle,\\
    |\phi_{\alpha,1}\rangle = \sqrt{1-y}|H\rangle - (-1)^{\alpha}\sqrt{y}|V\rangle,
    \end{aligned}
    \end{equation}
    where $|H\rangle$ and $|V\rangle$ represent the horizontal and vertical polarization states, respectively, and sends it to Bob.

\begin{figure}[b]
\includegraphics[scale=0.45]{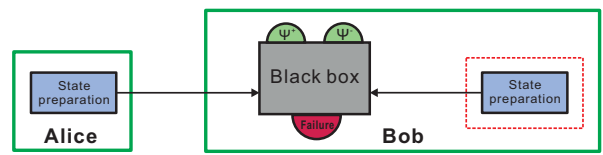}% Here is how to import EPS art
\caption{\label{fig:mdiQCT} (Color online) Schematic diagram of MDI-QCT. Here, Alice and Bob prepare, e.g., polarization states as Eq. (\ref{eq:states}) randomly using single-photon sources. Bob performs the BSM through the black box, which may be fabricated by Alice. The three outputs $|\Psi^{+}\rangle$, $|\Psi^{-}\rangle$ and "failure" of the black box represent the results of the BSM. If the output "failure" clicks, Bob will request Alice to restart the protocol. The red dotted box surrounding Bob's source implies that the black box cannot obtain the classical information about Bob's state preparation.}
\end{figure}

2. Bob prepares the state $|\phi_{\beta, b}\rangle$ ($\beta\in\{0, 1\},b\in\{0, 1\}$) randomly and independently of Alice. Then, he inputs Alice's and his states into a black box to perform a Bell state measurement (BSM) that projects $|\phi_{\alpha, a}\rangle$ and $|\phi_{\beta, b}\rangle$ into a Bell state (see Fig.~\ref{fig:mdiQCT}). If the BSM does not obtain a Bell state, then the black box outputs a "failure" click. Bob now will ask Alice to restart step 1. Otherwise, Bob records the corresponding result. Bob needs only to identify Bell states $|\Psi^{\pm}\rangle = \frac{1}{\sqrt{2}}(|01\rangle \pm |10\rangle)$, implying the black box can only use linear optical elements. Theoretical probabilities of the outcomes $|\Psi^{\pm}\rangle$ for different combinations of $|\phi_{\alpha, a}\rangle$ and $|\phi_{\beta, b}\rangle$ are shown in Table \ref{tab:Bell states}.

3. Bob sends Alice a random bit $b' \in\{0, 1\}$.

4. Alice reveals $\{\alpha, a\}$.

5. Bob compares $\{\alpha, a\}$ and $\{\beta, b\}$ with the cells in Table \ref{tab:Bell states}.
If the combination of $\{\alpha, a\}$ and $\{\beta, b\}$ corresponds to the cell with probability 0, then Bob detects Alice cheating and aborts the protocol. Otherwise, the outcome of the coin value is $x = a \oplus b'$.

Furthermore, because Bob does not need to protect the BSM from unwanted leakage of classical information to the outside, we can permit an agent of Bob performing the BSM in the middle of Alice and Bob, and the secure transmission distance of MDI-QCT will be doubled.

The reasons that MDI-QCT is loss tolerant with single-photon source are explained as follows. First, MDI-QCT is a bit-commitment-based QCT protocol, and Alice partially commits to either $a=0$ or $a=1$ at step 1. The density matrices that correspond to Alice's partial commitments are the same with BBBG09. Thus, as proved in \cite{berlin2009fair}, Bob cannot through the UD or MCD to obtain a better discrimination of Alice's commitment than the MED.

Second, since Alice can fabricate the measurement device, the black box has the chance to discriminate the four honest states randomly sent by Bob and announce the result to Alice. The discrimination processes of the four states can be regarded as the following form mathematically. Bob sends the black box a photon which is prepared with the same \emph{prior} probability $\frac{1}{4}$ in one of the 4 given honest states. Thus, the density operator of the state received by the box is
\begin{eqnarray}\label{eq:density}
\rho=\frac{1}{4}\displaystyle\sum_{i=1}^{4}\rho_{i}=\frac{I}{2},
\end{eqnarray}
where $I$ is the identity operator in two-dimensional Hilbert space, $\rho_{1}=|\phi_{0,0}\rangle\langle\phi_{0,0}|, \rho_{2}=|\phi_{0,1}\rangle\langle\phi_{0,1}|, \rho_{3}=|\phi_{1,0}\rangle\langle\phi_{1,0}|$, and $\rho_{4}=|\phi_{1,1}\rangle\langle\phi_{1,1}|$. Discriminating multiple states is a nontrivial task. However, the four honest states are linearly dependent, which implies that the UD does not apply here \cite{chefles1998unambiguous}. On the other hand, Eq.~(\ref{eq:density}) implies that the the identity operator can be resolved as a weighted sum over the density operators $\rho_{i}$. It follows from the Ref.~\cite{herzog2012optimal} that the MCD of the four honest states can be done without inconclusive results, namely, the MCD coincides with the MED and the maximal guessing probability can not be increased by admitting inconclusive results. Therefore, neither Bob nor Alice could increase his/her \emph{bias} by UD and MCD due to the inevitable losses in practical implementation.

\begin{table*}
\caption{\label{tab:Bell states}Theoretical probabilities of Bell states $|\Psi^{\pm}\rangle$ for different combinations of $|\phi_{\alpha, a}\rangle$ and $|\phi_{\beta, b}\rangle$. These can be calculated by the interferences of the honest states at the beam splitter.}
\begin{ruledtabular}
\begin{tabular}{cccccccccc}
 &\multicolumn{4}{c}{$|\Psi^{+}\rangle$}&\multicolumn{5}{c}{$|\Psi^{-}\rangle$}\\
 (a)&$|\phi_{0,0}\rangle$&$|\phi_{0,1}\rangle$&$|\phi_{1,0}\rangle$
&$|\phi_{1,1}\rangle$&(b)&$|\phi_{0,0}\rangle$&$|\phi_{0,1}\rangle$&$|\phi_{1,0}\rangle$
&$|\phi_{1,1}\rangle$\\ \hline

 $|\phi_{0,0}\rangle$&$2y(1-y)$&$\frac{1}{2}(1-2y)^{2}$ &$0$&$\frac{1}{2}$ & $|\phi_{0,0}\rangle$&$0$&$\frac{1}{2}$ &$2y(1-y)$&$\frac{1}{2}(2y-1)^{2}$\\

 $|\phi_{0,1}\rangle$&$\frac{1}{2}(1-2y)^{2}$&$2y(1-y)$&$\frac{1}{2}$&$0$ & $|\phi_{0,1}\rangle$&$\frac{1}{2}$&$0$&$\frac{1}{2}(2y-1)^{2}$&$2y(1-y)$\\

 $|\phi_{1,0}\rangle$&$0$&$\frac{1}{2}$&$2y(1-y)$&$\frac{1}{2}(2y-1)^{2}$ & $|\phi_{1,0}\rangle$&$2y(1-y)$&$\frac{1}{2}(1-2y)^{2}$&$0$&$\frac{1}{2}$\\

 $|\phi_{1,1}\rangle$&$\frac{1}{2}$&$0$&$\frac{1}{2}(2y-1)^{2}$&$2y(1-y)$ & $|\phi_{1,1}\rangle$&$\frac{1}{2}(1-2y)^{2}$&$2y(1-y)$&$\frac{1}{2}$&$0$\\
\end{tabular}
\end{ruledtabular}
\end{table*}

\section{Security of MDI-QCT} \label{security}

In this section, we discuss the correctness and security of the proposed protocol. For the correctness, we estimate the honest abort probability. For the security, we calculate each party's maximal cheating probability and the coefficients of the states in Eq.~(\ref{eq:states}) to make the protocol \emph{fair}.

\subsection{Honest abort probability} \label{h}

\begin{figure}[b]
\includegraphics[scale=0.45]{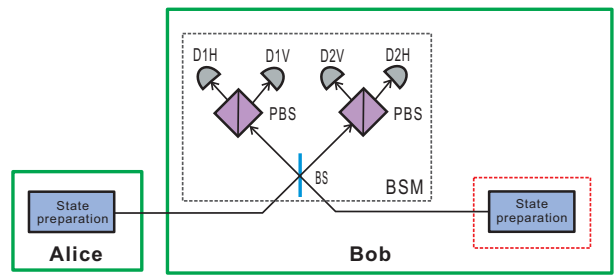}% Here is how to import EPS art
\caption{\label{fig:honestmdiQCT} (Color online) Schematic diagram of MDI-QCT when both parties are honest. Here, BS stands for 50:50 beam splitter and PBS stands for polarization beam splitter. Alice and Bob prepare polarization states as Eq.~ (\ref{eq:states}) randomly using single-photon sources. Bob performs the BSM and records the results for the verification of Alice's honesty. A joint click on $D1H$ and $D2V$, or $D1V$ and $D2H$, represents a projection into Bell state $|\Psi^{-}\rangle$. And a joint click on $D1H$ and $D1V$, or $D2H$ and $D2V$, represents a projection into Bell state $|\Psi^{+}\rangle$.}
\end{figure}

Like practical QKD systems, MDI-QCT inevitably will suffer from imperfections in the real world. As suggested in \cite{hanggi2011tight}, to make the protocol allow some amount of noise in the state preparation, transmission and detection processes, the players will abort with some probability when both of them are honest. For MDI-QCT is loss-tolerant when both parties use single-photon state sources, there is always a successful BSM result. In its simplest case, we also have required that Alice and Bob perfectly characterize their encoded quantum states. And we could attribute the noise in transmission lines to the dark counts of the detectors. With these simplification, the honest parties will abort in the situation that the first successful BSM is due to dark counts. Because both parties now are honest, we can describe the BSM specifically as shown in Fig.~\ref{fig:honestmdiQCT}. If the projection of BSM is $|\Psi^{+}\rangle$, Bob will abort with probability $\frac{1}{4}$, since Alice and Bob prepare the states randomly and there are four cells with probability 0 in Table \ref{tab:Bell states} for $|\Psi^{+}\rangle$. The abort probability is same for projection $|\Psi^{-}\rangle$. Here, we assume the four single-photon detectors in the BSM have identical properties. The total honest abort probability is

\begin{eqnarray}
Pr_{H} = &&2\cdot \frac{1}{4}[(1-t_{A})(1-t_{B})\cdot 2d^{2} + t_{A}(1-t_{B})\eta \cdot d +\nonumber\\
 &&t_{B}(1-t_{A})\eta \cdot d + t_{A}(1-t_{B})(1-\eta) \cdot 2d^{2} +\nonumber\\
&&t_{B}(1-t_{A})(1-\eta) \cdot 2d^{2} + t_{A}t_{B}(1-\eta)^{2}\cdot 2d^{2}]
\end{eqnarray}
where $d$ is the dark count rate and $\eta$ is the quantum efficiency of the detector; $t_{A}=10^{-0.02l_{A}}$ and $t_{B}=10^{-0.02l_{B}}$ are the $l_{A}$ and $l_{B}$ lengths ($km$) channel transmittance of Alice's and Bob's fiber, respectively; here, we let the loss coefficient in the fiber be $0.2dB/km$. Note, we have omitted the factor $(1-d)^{2}\sim 1$, which implies that the other two detectors do not click in order to obtain a successful projection.

\begin{figure}[b]
\includegraphics[scale=0.6]{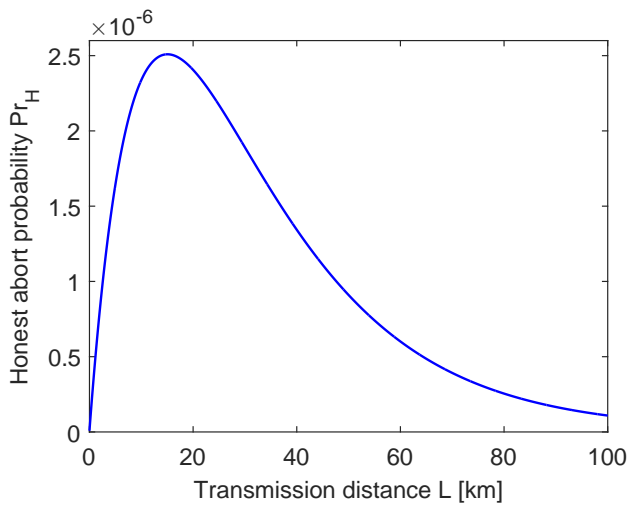}% Here is how to import EPS art
\caption{\label{fig:prh} (Color online) Relationship between the honest abort probability and the transmission distance. The quantum efficiency $\eta$ of the detector is made $10\%$. The dark count rate $d$ is $10^{-4}$, which is higher than the usual case because we have attributed the noise in transmission lines to dark counts. For simplification, we make $l_{A}=l_{B}=L$.}
\end{figure}

The relationship between the honest abort probability and the transmission distance is plotted in Fig. \ref{fig:prh}. At a shorter distance, less than $20 km$ in the figure, the case that a photon and a dark counts lead to a BSM result is more likely. However, at a longer distance, the probability that a BSM projection is completely due to dark counts increases.  From Fig. \ref{fig:prh} we can see that the coincidence detections of the BSM make our protocol more robust against the detector dark counts.

\subsection{Dishonest Bob's maximal cheating probability} \label{pb}
Assume that Bob tries to bias the coin towards the value $x=0$ (the analysis for $x=1$ is similar because the protocol is symmetric). We also suppose that the conditions are all in favor of a dishonest Bob, i.e., the channels are perfect and Bob's detectors are perfect. These only maximize Bob's cheating probability.

The optimal cheating strategy of a malicious Bob is to perform an optimal measurement to discriminate the received states sent by honest Alice, i.e., to obtain $a$. Then, Bob can decide the value of random bit $b'$ in step 3 to obtain his desired coin's outcome.
As the honest states sent by Alice are no different in form from those in BBBG09 \cite{berlin2009fair}, dishonest Bob's optimal cheating strategy is the same as theirs. For the case where honest Alice uses a single-photon source, Bob's maximal cheating probability is $Pr_{B}[x]=y$ \cite{berlin2009fair}.

In the above cheating strategy, a dishonest Bob will deviate the MDI-protocol to obtain a maximal cheating probability. However, our protocol is designed only to protect an honest Bob, thus it is no matter that the advantage of the MDI paradigm does not exists for a dishonest Bob.

\subsection{Dishonest Alice's maximal cheating probability} \label{pa}
Assume that Alice tries to bias the coin towards the value $x=0$ (the analysis for $x=1$ is similar because the protocol is symmetric). In the following, we assume that Alice builds the measurement device which contains her cheating device, and gives it to Bob. Bob only needs to protect the classical information of his state preparation from leakage to the untrusted measurement device and Alice. Thus, the black box cannot access Bob's classical information used to prepare quantum states. On the other hand, the outcome of the BSM cannot reveal Bob's quantum state with certainty. Therefore, we can let the black box send classical information to Alice, which can be simply done by classical communication device implanted into the black box beforehand. We also assume that the detectors and both parties' single-photon sources are perfect, and the channel is lossless. All these will maximize Alice's cheating probability. Therefore, the protocol can be modified as follows. First, Bob randomly sends the state $|\phi_{\beta, b}\rangle$ to the measurement black box. Then the black box does the cheating strategy of Alice and announces a result of $|\Psi^{+}\rangle$, $|\Psi^{-}\rangle$ or failure to both parties correspondingly. The black box can also send other classical information to Alice. The remaining steps are the same with steps 3, 4 and 5 in the honest MDI-QCT.

Based on the above modified protocol, we first introduce an individual attack where Alice places a measurement device to discriminate Bob's state in the black box. Now, Alice does not send any state in step 2. The box will send the discrimination result to Alice and output a BSM projection randomly to both parties. Once the box correctly discriminates Bob's state, then after step 3, Alice can reveal appropriate $\{\alpha, a\}$ according to Table \ref{tab:Bell states} and $b'$ to obtain her desired coin's value without being caught cheating. The four states randomly sent by honest Bob are pairs of two orthogonal states with the same \emph{prior} probability $\frac{1}{4}$. As proven in Ref. \cite{bae2013structure}, the guessing probability of the MED is $\frac{1}{2}$. On the other hand, her discrimination has an error rate of $\frac{1}{2}$. As from her point of view, Alice can not determine whether the result of the MED is wrong. Thus, in this time, the probability of her choice of $\{\alpha, a\}$ passing Bob's test is $\frac{1}{2}$, for there are two states $|\phi_{\beta, b}\rangle$ in Table \ref{tab:Bell states} that will make Alice pass the verification and each has a probability $\frac{1}{4}$. To sum up, she can bias the coin in the individual attack with probability $Pr_{A}^{ind}[x]=\frac{1}{2}+\frac{1}{2}\cdot \frac{1}{2}=\frac{3}{4}$.

In the following we discuss a coherent attack. The success of Alice's cheating should satisfy two conditions. One is that the combination of the result of the BSM and their states must pass Bob's verification at the last step. The other is that the coin's value is her desired one. Note that the second one can always be satisfied because in the protocol Alice will announce her state only after Bob has told a random bit $b'$. That is to say, Alice only has to try her best to pass Bob's verification via the operations in the measurement black box. Here, we assume Alice always let the black box do BSM. In the following, we calculate Alice's maximal cheating probability under a coherent attack with this assumption.

Considering the MLC \cite{lo1998quantum,mayers1997unconditionally,lo1997quantum,lo1997quantum} no-go theorem, the general optimal cheating strategy for a dishonest Alice is an optimal coherent attack. Specifically, Alice submits to Bob one half of an entangled state, which will maximize her probability of passing Bob's test. However, because MDI-QCT is a bit-commitment-based QCT protocol, the pioneering work of Spekkens and Rudolph \cite{spekkens2002optimization}, which gives a thorough investigation of the optimal coherent attack on bit-commitment-based quantum cryptography tasks, applies here.

In MDI-QCT, two honest pure states $|\phi_{\alpha, 0}\rangle$ correspond to Alice's committed bit 0 and two honest pure states $|\phi_{\alpha, 1}\rangle$ correspond to Alice's committed bit 1. This configuration agrees with case 1 of section 6.4.1 in \cite{spekkens2002optimization}.
Hence, there are three optimal states for Alice to submit to Bob in step 1 of the protocol: the first is $|+\rangle$, which has equal trace distance \cite{nielsen2010quantum} with $|\phi_{0,0}\rangle$ and $|\phi_{1,1}\rangle$; the second is $|-\rangle$, which has equal trace distance with $|\phi_{0,1}\rangle$ and $|\phi_{1,0}\rangle$; and the third is any mixture of the states $|+\rangle$ and $|-\rangle$, where $|\pm\rangle=\frac{1}{\sqrt{2}}(|0\rangle \pm|1\rangle)$. Now, Alice does not need prepare an entangled state.
Because all the optimal states lead to a same maximal cheating probability, here we take the state $|+\rangle$ for illustration. The probabilities for states $|+\rangle$ (or $|-\rangle$) and $|\phi_{\beta, b}\rangle$ being projected into Bell states $|\Psi^{\pm}\rangle$ are shown in Table \ref{tab:Cheating Bell states}.

In the following, we calculate the maximal probability that Alice can force the coin towards 0 with optimal state $|+\rangle$. We will suppose that the conditions are all in favor of a dishonest Alice, i.e., the channels and the detectors are perfect and the classical information of the BSM is leaked to herself. After step 3, Alice will reveal state $(\alpha=b',a=b')$ to Bob. Now, we should find the probability that the combinations of $(\alpha=b',a=b')$ and $(\beta, b)$ fall into the cells corresponding to probability 0 in Table \ref{tab:Bell states}, i.e., only when Alice will be caught cheating. There are four cases of the combinations: the first two are $\{(\alpha=0,a=0),(\beta=1, b=0)\}$ and $\{(\alpha=1,a=1),(\beta=0, b=1)\}$ for outcome $|\Psi^{+}\rangle$; the last two are $\{(\alpha=0,a=0),(\beta=0, b=0)\}$ and $\{(\alpha=1,a=1),(\beta=1, b=1)\}$ for outcome $|\Psi^{-}\rangle$. As shown in Table \ref{tab:Cheating Bell states}, the probabilities of the four cases, where Alice's states are replaced by the \emph{real} sent state $|+\rangle$, are all $\frac{1-2\sqrt{y(1-y)}}{2}$.
Because Bob prepares his states randomly and independently of Alice, and he selects $b'$ randomly, Alice can bias the coin with probability:
\begin{eqnarray}
Pr_{A}^{coh}[x] = &&1 -  \frac{1}{2}\cdot \frac{1}{4} \cdot \frac{1-2\sqrt{y(1-y)}}{2}\cdot2-\nonumber\\
 &&\frac{1}{2}\cdot \frac{1}{4} \cdot \frac{1-2\sqrt{y(1-y)}}{2}\cdot2\nonumber\\
= &&\frac{3+2\sqrt{y(1-y)}}{4},
\end{eqnarray}
which is same as with BBBG09 \cite{berlin2009fair}.

At last, we calculate the coefficients of the states in Eq.~(\ref{eq:states}) to make the protocol \emph{fair}. For the purpose of fairness, let $Pr_{A}^{coh}[x] = Pr_{B}[x]$; thus, we have $y = 0.9$ for the state preparation and basis choice processes of the MDI-QCT protocol when each party uses a single-photon source. We can calculate the values
\[\varepsilon_{A}=\varepsilon_{B}=0.4,\]
which implies the \emph{bias} of loss-tolerant MDI-QCT is 0.4 under this coherent attack. This value is the same with BBBG09.

\begin{table*}
\caption{\label{tab:Cheating Bell states}Theoretical probabilities of Bell states $|\Psi^{\pm}\rangle$ for Alice's coherent cheating strategy. The probabilities have been normalized, for the result of BSM is either $|\Psi^{+}\rangle$ or $|\Psi^{-}\rangle$ if the step 2 succeeds.}
\begin{ruledtabular}
\begin{tabular}{cccccccccc}
 &\multicolumn{4}{c}{$|\Psi^{+}\rangle$}&\multicolumn{5}{c}{$|\Psi^{-}\rangle$}\\
 (a)&$|\phi_{0,0}\rangle$&$|\phi_{0,1}\rangle$&$|\phi_{1,0}\rangle$
&$|\phi_{1,1}\rangle$&(b)&$|\phi_{0,0}\rangle$&$|\phi_{0,1}\rangle$&$|\phi_{1,0}\rangle$
&$|\phi_{1,1}\rangle$\\ \hline

 $|+\rangle$&$\frac{1+2\sqrt{y(1-y)}}{2}$&$\frac{1-2\sqrt{y(1-y)}}{2}$&$\frac{1-2\sqrt{y(1-y)}}{2}$&$\frac{1+2\sqrt{y(1-y)}}{2}$ &$|+\rangle$&$\frac{1-2\sqrt{y(1-y)}}{2}$&$\frac{1+2\sqrt{y(1-y)}}{2}$&$\frac{1+2\sqrt{y(1-y)}}{2}$&$\frac{1-2\sqrt{y(1-y)}}{2}$\\

$|-\rangle$&$\frac{1-2\sqrt{y(1-y)}}{2}$&$\frac{1+2\sqrt{y(1-y)}}{2}$&$\frac{1+2\sqrt{y(1-y)}}{2}$&$\frac{1-2\sqrt{y(1-y)}}{2}$ &$|-\rangle$&$\frac{1+2\sqrt{y(1-y)}}{2}$&$\frac{1-2\sqrt{y(1-y)}}{2}$&$\frac{1-2\sqrt{y(1-y)}}{2}$&$\frac{1+2\sqrt{y(1-y)}}{2}$\\

\end{tabular}
\end{ruledtabular}
\end{table*}

\section{MDI-QCT for weak coherent state source} \label{wcp}
In practical quantum cryptography system, because a perfect single-photon source is beyond the state-of-the-art technology, a weak coherent state source is widely used. To implement the MDI-QCT protocol with a weak coherent state source and avoid the multi-photon security loophole as noted in \cite{berlin2009fair}, we can utilize the technique in PCDK11. Namely, Alice sends a fixed number $K$ of pulses in step 1, where $K$ is determined by the security analysis. If none of the BSM in step 2 is successful, Bob aborts. Otherwise, let $j$ label the first identified Bell state, which Bob will tell Alice in step 3. The rest of the protocol remains the same, except that the values of $\{\alpha, a, \beta,  b\}$ are all labeled by the index $j$. Obviously, using this technique comes with a price, i.e., the protocol now is not completely impervious to losses.

The security of MDI-QCT with weak coherent state source is the same with its single-photon source counterpart when Alice is dishonest. As for a malicious Bob, his maximal cheating probability is changed because of the multi-photon pulses emitted by the weak coherent state source. Fortunately, the calculation has been done in \cite{pappa2011practical}. Therefore, one can use the result in \cite{pappa2011practical} to obtain the value of $y$ and $K$.

\section{Conclusion}

In conclusion, some side channel (including source \cite{PhysRevA.91.032326} and detector) attacks presented in practical QKD can also arise in other practical quantum cryptographic tasks with mistrustful parties. In this paper, we have proposed MDI-QCT to remove all the detector side-channels in practical QCT. Our results suggest that the idea in MDI-QKD can also be applied to QCT, as well as other quantum cryptographic tasks based on QCT, e.g., two-party secure computation. Thus we extend the scope of the skill of MDI-QKD into a very rich field. Furthermore, our protocol can increase the transmission distance if we permit an agent of Bob to perform the BSM in the middle of  Alice and Bob.

We have already proven the security of MDI-QCT against a dishonest Alice under a coherent attack, which is probably the optimal coherent attack because of the work of Ref. \cite{spekkens2002optimization}. It is interesting to have a more general security analysis in the future. Another theoretic research needed is to combine the source flaws into security analysis. Because in MDI-QCT, we have assumed that Alice's and Bob's sources can be trusted. It should examine this condition carefully in practice.

In the experiments, MDI-QCT can be implemented with a weak coherent state source. The platforms for the implementation of MDI-QKD \cite{rubenok2013real,da2013proof,liu2013experimental,xu2013long,tang2014experimental,tang2014measurement} can also be used here with several modifications. Thus, we expect MDI-QCT to be widely utilized in practical QCT systems in the future.

\begin{acknowledgments}
The authors would like to thank the anonymous referees, they had made a great deal of constructive comments to improve the quality of this article. This work was supported by the National Basic Research Program of China (Grants No. 2011CBA00200 and No. 2011CB921200), the National Natural Science Foundation of China (Grant Nos. 61475148, 61201239, 61205118, 11304397), and the "Strategic Priority Research Program (B)" of the Chinese Academy of Sciences (Grant Nos. XDB01030100, XDB01030300).

\end{acknowledgments}

%\appendix

%\section{Appendixes}

\bibliography{mdiQCT}% Produces the bibliography via BibTeX.

\end{document}